\documentclass[pra,twocolumn,showpacs,superscriptaddress,floatfix]{revtex4}
\usepackage{epsfig}
\usepackage{amsmath}%
\usepackage{amsfonts}%
\usepackage{amssymb}%
\usepackage{graphicx}

\newcommand\br{\mathbf{r}}
\newcommand\bk{\mathbf{k}}
\newcommand\kp{\mathbf{k}_\parallel}
\newcommand\rd{\mathrm{d}}
\newcommand\bff{\mathbf{f}}

\begin{document}

\title{Casimir-Polder interaction between an atom and a dielectric slab}
\author{Ana Mar\'{\i}a Contreras Reyes and Claudia Eberlein}
\affiliation{Dept of Physics \& Astronomy, University of Sussex,
  Falmer, Brighton BN1 9QH, England}  
\date{\today}
\begin{abstract} We present an explicit analytic calculation of the 
energy-level shift of an atom in front of a non-dispersive and
non-dissipative dielectric slab. We work with the fully quantized
electromagnetic field, taking retardation into account. We give the shift as
a two-dimensional integral and use asymptotic analysis to find expressions
for it in various retarded and non-retarded limiting cases. The results can
be used to estimate the energy shift of an atom close to layered
microstructures.
\end{abstract}
 \maketitle

\section{Introduction}

Control and manipulation of cold atoms have become fundamentally important
due to their central role in the development of nanotechnology and as a tool
for investigating the mechanisms underlying macroscopic manifestations of
quantum physics \cite{Chan}. It now seems feasible to control them on
a $\mu$m length-scale by utilizing microstructured surfaces --- also
known as atoms chips --- with promising areas of application such as
quantum information processing with neutral atoms, integrated atom
optics, precision force sensing, and studies of the interaction
between atoms and surfaces.  

For this reason, e.g., experiments using Bose-Einstein condensates for
measuring the Casimir-Polder force \cite{CP} have been developed. Typically,
the dielectric substrate utilized in such experiments \cite{Obrecht} carries
a very thin top layer of another material, generally graphite or
gold. However, in explicit analytic theories, this finite thickness has
often been neglected and the system has been treated as a semi-infinite
half-space. Here we are aiming at an approach that lets us include the
thickness of such a layer as a parameter into the calculation, allowing us
to obtain analytic expressions for the Casimir-Polder force on an atom.

We are going to consider a ground-state atom close to a non-dispersive
dielectric slab, which is one of the fews systems of high symmetry for that
the electromagnetic field can be quantized through an exact normal-mode
expansion with manageable effort \cite{Khosravi}. The assumption of absent
dispersion and absorption is working well for all but very few systems,
namely those where the atom has a strong transition very near an absorption
line in the medium, which in practice is something very difficult to
engineer. Nevertheless, if one wishes to consider more complex systems with
atoms near absorbing boundaries, the calculation will require other methods
to study quantum electrodynamics \cite{Abrikosov}. However, applying such
results to a particular case might require extensive numerical calculations
for the case at hand. Using those types of formalism, the energy-level shift
of atoms due to the presence of media with diverse magneto-electric
properties has been calculated for several systems \cite{Buhmann}.  Here, by
contrast, we are not aiming at general expressions.  The focal point of the
present work is to obtain simple and practical formulae that are useful for
estimates and can be applied very easily to experimental situations.

The energy shift in an atom close to a dielectric slab comes about due to
its interaction with electromagnetic field fluctuations, which in turn are
affected by the presence of the slab. Thus, a quantization of the
electromagnetic field in the presence of a layered system is required. Even
though the field quantization for this system had been studied previously
\cite{Khosravi}, we have recently re-considered the problem and provided a
proof of the completeness of the electromagnetic field modes that was
missing in previous works on the problem. This proof of completeness is very
useful in that it removes any ambiguity in how to normalize and sum over the
electromagnetic field modes, and in this way also establishes the correct
density of states which had previously been a subject of disagreement
\cite{Zakowicz}.

By solving the Helmholtz equation and imposing the corresponding continuity
conditions at the faces of the slab, it was shown in \cite{Khosravi,
completeness} that the field modes for this system comprise of travelling
and trapped modes. The travelling modes have a continuous frequency
spectrum, and are composed of incident, reflected and transmitted parts
outside the slab. The trapped modes arise due to solutions of the
Helmholtz equation with purely imaginary normal wave vector outside the
slab. Physically, they come about due to repeated total internal reflection
inside the dielectric, and emerge as evanescent fields outside the
slab. They exist only at certain discrete frequencies which depend on
polarization direction and parity and are obtained through the dispersion
relations.
  
The atomic energy shift is obtained by means of second-order perturbation
theory, and involves a product of electromagnetic mode functions which is
summed over intermediate virtual photon states. Thus the atomic energy shift
receives two quite separate contributions: one from the continuous set of
travelling modes, and the other from the discrete set of trapped modes. The
first is an integral over wave numbers, and the second a sum over discrete
wave numbers that satisfy a quite complicated dispersion relation. In
practice, both must be considered together at all times to avoid divergent
terms appearing in each separate contribution but cancelling between them.
This technically seemingly hopeless task can, however, be dealt with simply
and elegantly by using the summation method of \cite{completeness}, which
re-expresses both the integral over continuous  modes and the sum over
discrete modes as a single contour integral in the complex plane.
We shall show below that this trick yields a closed-form expression for the
atomic energy shift and permits easy asymptotic analysis for various
regimes, yielding the kinds of simple formulae that we are after for
estimating the effect of the layer thickness in experimental situations.

\section{Description of the system}
\begin{figure}[t]
\vspace*{-1mm}
\begin{center}
\centerline{\epsfig{figure=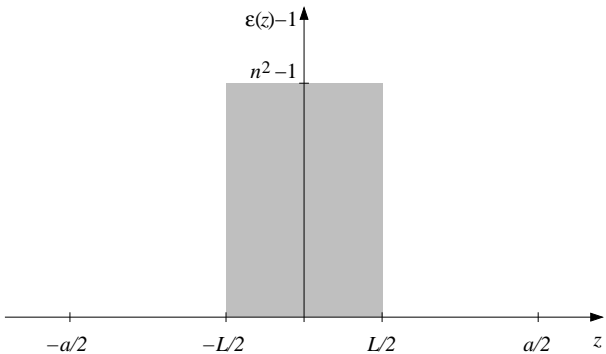,
  width= 8.0cm}}
\end{center}
\vspace*{-5mm} \caption{The geometry of the dielectric
  slab.} 
\label{system} 
\end{figure}

We consider a dielectric slab of finite thickness $L$ surrounded by vacuum,
as is shown in Fig. \ref{system}.  We assume the material to be a
non-dispersive and non-absorbing dielectric, which is a simple but good
model for an imperfectly reflecting material. Thus the material is
characterized solely by its refractive index $n$, which is real and the same
for all frequencies.  While any real material has of course to be
transparent at infinite frequencies, this non-dispersive model captures the
essential properties of an imperfect reflector. In particular, it includes
evanescent waves, whose absence in perfect-reflector models can be
problematic \cite{robaschik}.

Since the dielectric is homogeneous in the $x$ and $y$ directions, the 
dielectric permittivity of the configuration depends only on the $z$ 
coordinate and is given by  
\begin{displaymath}
\varepsilon(z) = \left \{
\begin{array}{ll}
n^2  & \qquad \textrm{for} \qquad -L/2 \leq z \leq L/2\;, \\
1 & \qquad \textrm{for} \qquad |z| \geq L/2\;.
\end{array} \right.
\end{displaymath}

We assume the atom to be neutral and in its ground state. Also, we shall
make use of the electric dipole approximation, which is adequate because,
for the relevant modes, the electromagnetic field varies slowly over the
size of the atom. We assume that the atom's center is fixed at the position
${\bf r}_0=(0,0,z_0)$.

The model assumes that the interaction between the atom
and the surface is purely electromagnetic, i.e. that there is negligible
wave-function overlap between the atomic electron and the surface.

We shall work with an interaction Hamiltonian between the atom and the
quantized electromagnetic field that is given by
\begin{equation}\label{intHamiltonian}
H_{\mbox{\footnotesize int}}=-\boldsymbol{\mu} \cdot {\bf E}({\bf r}, t)\;,
\end{equation}
which is the lowest-order multipole Hamiltonian and corresponds to the
electric-dipole interaction. In this equation, $\boldsymbol{\mu} = e({\bf
r}-{\bf r}_0)$ is the electric-dipole moment of the atomic electron, and
${\bf E}({\bf r}, t)$ is the transverse electric field. Unlike the
minimal-coupling Hamiltonian ${\bf p} \cdot {\bf A}$, the Hamiltonian
(\ref{intHamiltonian}) includes the electrostatic interaction between the
atomic dipole and its images on the other side of vacuum-dielectric
interfaces. As shown previously in a similar context \cite{Eberlein}, the
Hamiltonian (\ref{intHamiltonian}) may be more convenient for calculations
that aim to derive energy shifts in cases where the retardation of the
electromagnetic interaction matters.

The quantization of the electromagnetic field has been discussed in detail
previously \cite{Khosravi,completeness}, and thus, we shall only sketch
the procedure here. We work with the electromagnetic
potentials $\Phi (\br, t)$ and ${\bf A}(\br, t)$ and choose the
generalized Coulomb gauge
\begin{equation}\label{gauge}
{\nabla}\cdot \left[\varepsilon(z){\bf A}(\br)\right] =0. 
\end{equation}
Furthermore, since the overall system is neutral, we can set $\Phi (\br,
t)=0$. Thus the field equations reduce to the wave equation for ${\bf
A}(\br, t)$ everywhere except right on the interfaces $z=\pm L/2$.  At the
interfaces we solve Maxwell's equations directly by imposing the
continuity conditions
\begin{equation}\label{ContCond}
 E_{\|}, \quad D_\perp, \quad {\bf B} \qquad {\rm continuous}\;.
\end{equation}
In this way the electromagnetic field modes can be written, for travelling
modes $\bff_{\nu}^{L,R}$ as left- or right-incident waves made up of
incoming, reflected and transmitted parts, and for trapped modes
$\bff_{\nu}^{S,A}$, as symmetric or antisymmetric waves inside the slab with
evanescent fields outside. We list these modes in Appendix \ref{app:modes}.  

Equipped with a complete set of solutions to the classical field equations,
we can proceed to quantize the electromagnetic field by using the technique
of canonical quantization, i.e. by introducing annihilation and creation
operators $a_{\nu}$, $a_{\nu}^{\dagger}$. Then the expansion for the
electric field operator ${\bf E} (\br, t)$ in terms of the normal modes
$\bff_{\nu}(\br)$ reads
\begin{equation} \label{E}
{\bf E} (\br, t)= {\rm i} \sum_{\nu} \sqrt{\frac{\omega_{\nu}}{2
\varepsilon_0}} \left[ 
a_{\nu} e^{-{\rm i} \omega_{\nu} t} \bff_{\nu}(\br) - a_{\nu}^
{\dagger} e^{{\rm i} \omega_{\nu} t} \bff_{\nu}^{*}(\br) \right].
\end{equation} 
where the subscript $\nu = (\bk, \lambda)$ is a composite label including
both the polarization $\lambda=\mbox{TE}, \mbox{TM}$ and the
wave vector $\bk$.

\section{Energy-level shift}
Since the interaction Hamiltonian (\ref{intHamiltonian}) is linear in the
electric field, whose vacuum expectation value vanishes, there is no energy
shift to first order in $H_{\mbox{\footnotesize int}}$. Therefore, the
lowest-order contribution to the shift comes from second-order perturbation
theory, so that the shift is of first order in the fine-structure constant
$\alpha$ \cite{fn1},
\begin{equation}
\Delta E= -\sum_{j \neq i} \sum_{\nu} \frac{\left| \left \langle j;
\nu |\boldsymbol{\mu} \cdot {\bf \hat E}(\br, t) |i;0 \right \rangle 
\right|^2} {E_j -E_i + \omega_{\nu}},
\end{equation}
where ${\bf {\hat E}}({\bf r}, t)$ is given by Eq.~(\ref{E}). In this
equation, the intermediate state $\left| j;\nu \right \rangle$ is a
composite state with an atom in the excited state $|j \rangle$ and the
electromagnetic field carrying a photon of energy $\omega_{\nu}$. Similarly,
the initial state $|i;0 \rangle$ describes an atom in its ground state $|i
\rangle$ and the electromagnetic field in the vacuum state. In the electric
dipole approximation we can write
\begin{equation}
\left \langle j| \bff(\br) \cdot \boldsymbol{\mu} |i \right \rangle \simeq
\bff(\br_0) \cdot \left \langle j| \boldsymbol{\mu} |i \right \rangle,
\end{equation}
since the field varies slowly over the size of the atom and we can
therefore assume that across the atom it is almost the same as at its center 
${\bf r}_0=(0,0,z_0)$. With this, the shift reads
\begin{equation}
\Delta E = 
 - \frac{1}{2 \varepsilon_0} \sum_{j \neq i} \sum_{\nu} \frac{\omega_{\nu}} 
 {E_{ji}+\omega_{\nu}} \left|\bff^*_{\nu}(\br_0) \cdot \langle j | 
 \boldsymbol{\mu}| i\rangle \right|^2,
\end{equation} 
where we have introduced the abbreviation $E_{ji}=E_j-E_i$. 
Since $|i\rangle$ is a state of definite angular momentum, different
components of $\boldsymbol{\mu}$ lead to different intermediate states
$|j\rangle$ that are mutually orthogonal, and the shift simplifies
to
\begin{equation}
\Delta E = 
 - \frac{1}{2 \varepsilon_0} \sum_{j \neq i} \sum_{\nu} \frac{\omega_{\nu}} 
 {E_{ji}+\omega_{\nu}} \left|\bff^*_{\nu}(\br_0)|^2\; |\langle j | 
 \boldsymbol{\mu}| i\rangle \right|^2 .
\label{generalshift}
\end{equation} 
Furthermore, we
are going to abbreviate the moduli squares of the matrix elements of the
dipole-momentum operator $\boldsymbol{\mu}$ between the initial state $i$
and the intermediate states $j$, and distinguish only the components
parallel and perpendicular to the slab,
\begin{eqnarray}
&&|\mu_\sigma|^2 \equiv \left| \left \langle j|
 \mu_{\sigma} | i \right\rangle \right|^2 \quad {\rm with} \quad
 \sigma = x, y, z\nonumber\\
&&|\boldsymbol{\mu}_{||}|^2 \equiv | \langle j|\mu_x| i\rangle|^2 + 
| \langle j| \mu_y|i\rangle|^2  \nonumber\\ 
&& |\boldsymbol{\mu}_{\perp}|^2 
\equiv |\langle j|\mu_z|i\rangle|^2\;. \nonumber
\end{eqnarray}

The sum over $\nu$ in Eq.~(\ref{generalshift}) is a sum over all field
modes, which, as explained earlier and easily seen from Appendix
\ref{app:modes}, comprise a continuous set of travelling modes and a
discrete set of trapped modes. The contribution from the travelling modes
gives rise to the shift
\begin{eqnarray}
&& \hspace*{-8mm} \Delta E^{{\rm trav}}= - \frac{1}{2 \varepsilon_0}
  \sum_{j \neq i} \sum_{\lambda=TE,TM} \sum_{\sigma} \int \rd^3 {\bf k}
  \frac{\omega} {E_{ji}+\omega} \nonumber \\ \label{trav}&& \times
  \left ( |{\bf f}^L_{\bk\lambda\sigma} ({\bf r}_0)|^2 + 
  |{\bf f}^R_{\bk\lambda\sigma}({\bf r}_0)|^2 \right ) 
  \left| \mu_{\sigma} \right|^2,   
\end{eqnarray}
where the sum over $\sigma$ runs over the $x$, $y$ and $z$ components of the
dipole moment and of the polarization vector that is incorporated in the
mode functions. As we are interested in the change in the energy levels of
the atom solely due to the presence of the dielectric slab, we renormalize
the energy shift and remove from Eq. (\ref{trav}) the part that arises due
to the interaction between the atom and the electromagnetic field in free
space, i.e. the Lamb shift. Conveniently, this rids the calculation of any
divergences, provided travelling and trapped modes are considered together
(cf. e.g.~\cite{Wu-paper1}). The simplest way to implement this
renormalization of the shift is by subtracting the equivalent expression for
a transparent slab with $n=1$,
\begin{equation}
\delta E^{{\rm trav}} \equiv \Delta E^{{\rm trav}}
- \Delta E^{{\rm trav}}(n=1)\;.
\end{equation}
We decide to place the atom at a position $z_0>L/2$ to the right of the
slab, substitute the mode functions (\ref{fL},\ref{fR}), and get
\begin{eqnarray} \label{shift-trav}
&&  \hspace*{-5mm} \delta E^{{\rm trav}} = -\frac{1}{2 (2\pi)^3 
\varepsilon_0} \sum_{j \neq i} \sum_{\lambda, \,\sigma} \; \int \rd^3 
\bk \; \frac{\omega} {E_{ji}+ \omega}  \left|\mu_{\sigma} \right|^2 
\nonumber \\ && \times \; \hat e_{\lambda}^{\sigma} (\bk ^+) \hat 
e_{\lambda}^{\sigma *} (\bk ^-) \Big ( R_{\lambda} \; e^ {2{\rm i} 
k_z {z_0}} + R_{\lambda}^* \; e^ {-2{\rm i}k_z{z_0}}  \Big).  
\end{eqnarray}

Similarly, the contribution from the discrete set of trapped modes
reads 
\begin{eqnarray}
&& \hspace*{-15mm} \Delta E^{{\rm trap}}= - \frac{1}{2 \varepsilon_0}
  \sum_{j \neq i} \sum_{\lambda,\,\sigma} \sum_{k_z} 
  \int \rd^2 \kp \; 
  \frac{\omega} {E_{ji}+\omega} \nonumber \\
&& \times \; \left( | \bff^S_{\bk\lambda \sigma} (\br_0)|^2 + 
  |\bff^A _{\bk\lambda\sigma} ({\bf r}_0) |^2 \right ) 
  \left| \mu_{\sigma} \right|^2,  
\end{eqnarray}
which can be written in a more explicit form by substituting the
trapped modes to the right of the slab from Eq.~(\ref{fSA}),
\begin{eqnarray}\label{shift-trapp} 
&& \hspace*{-15mm} \delta E^{{\rm trap}} = -\frac{1}{2\varepsilon_0}
  \sum_{j \neq i} \sum_{\lambda,\,\sigma} \sum_{k_z}\int \rd^2 \kp \;
  \frac{\omega}{E_{ji} + \omega}  \left| \mu_{\sigma} \right|^2
  \nonumber \\ && \times \; \hat e_{\lambda}^{\sigma} (\bk ^+) \hat
  e_{\lambda}^{\sigma *} (\bk ^-) \left| M_{\lambda} \right|^2 \left|
  L_{\lambda}^{S,A} \right|^2 e^{-2\kappa {z_0}}.
\end{eqnarray}
We note that renormalization makes no difference to the trapped-modes
contribution to the shift, as the trapped modes vanish in the limit
$n\rightarrow 1$.

The total energy shift is obtained by combining the travelling-mode
contribution Eqs. (\ref{shift-trav}) and the trapped-mode contribution
(\ref{shift-trapp}). At first sight, this is very complicated, since the
former is given by an integral over $k_z$ while the latter involves a sum
over discrete values of $k_z={\rm i}\kappa$ that are solutions of the
dispersion relations (\ref{dispersion}). In addition, the shift
(\ref{shift-trav}) due to travelling modes and its counterpart
(\ref{shift-trapp}) due to trapped modes diverge when evaluated each on
their own, as observed before in similar circumstances \cite{Wu-paper1}.
What helps, is the observation that the reflection coefficients
$R_{\lambda}$ have poles in the complex $k_z$ plane at exactly the values of
$k_z={\rm i}\kappa$ that are solutions of the dispersion relations
(\ref{dispersion}). Furthermore, the residues around those poles are such
that the sum over $k_z$ in Eq.~(\ref{shift-trapp}) can be re-written as a
contour integral with the same integrand as in Eq.~(\ref{shift-trav}). Thus
the sum of Eqs.~(\ref{shift-trav}) and (\ref{shift-trapp}) can be combined
into a single contour integral in the complex $k_z$ plane.  In
Ref.~\cite{completeness} we have shown this to be the case in connection
with a proof of the completeness of the electromagnetic field modes around a
dielectric slab, and we refer the reader there for details. Using this
method to add Eqs.~(\ref{shift-trav}) and (\ref{shift-trapp}), we obtain for
the total energy shift
\begin{eqnarray} \label{shift}
&& \hspace*{-5mm} \delta E = -\frac{1}{2 (2\pi)^3 \varepsilon_0}
  \sum_{j \neq i} \sum_{\lambda, \sigma} \; \int \rd^2 \bk_{\|}
  \int_{\mathcal C} \rd k_z \; \frac{\omega}{E_{ji}+ \omega} \nonumber
  \\ && \times  \left| \mu_{\sigma} \right|^2 \hat e_{\lambda}^{\sigma}
  (\bk ^+) \hat e_{\lambda}^{\sigma *} (\bk ^-) R_{\lambda} (k_z,\kp)
  \; e^{2{\rm i}k_z{z_0}}\;, 
\end{eqnarray} 
with the integration path $\mathcal C$ as shown in Fig.~\ref{contour}. The
poles of $R_{\lambda}$ lie on the imaginary axis between 0 and 
${\rm i}\sqrt{n^2-1}k_\|/n$, so that $\mathcal C$ runs above them.
\begin{figure}
\begin{center}
{\includegraphics[width=0.5\textwidth]{contour.eps}}
\end{center}
\caption{\label{contour} \footnotesize{By closing the contour
    $\mathcal C$ one can choose a more suitable integration path
    $\mathcal C'$. In the figure, $\Gamma = \sqrt{n^2 -1}/n$.}}
\end{figure}
To manipulate this expression further, we sum over the two polarizations and
re-arrange the Cartesian components $\sigma =x,y,z$ into parallel and
perpendicular parts relative to the surface of the slab. The double integral
in $\bk_{\|}$ can be simplified by transforming into polar coordinates and
carrying out the integration in the azimuthal angle, so that the total
energy shift reads
\begin{equation}\label{deltaE-ret}
\delta E = -\frac{1}{2 \pi^2 \varepsilon_0} \sum_{j \neq i} 
\sum_{\sigma=\|,\perp} \; E_{ji}^3 \;S_{\sigma} \left|  
\boldsymbol{\mu}_{\sigma}  \right|^2 
\end{equation}
with parallel and perpendicular contributions given, respectively, by 
\begin{equation}
S_{\|} \equiv  \frac{1}{8 E_{ji}^3} \int_0^{\infty} \rd k_{\|} \; k_{\|}
I_{\|} \quad {\rm and} \quad S_{\perp} \equiv \frac{1} {4 E_{ji}^3}
\int_0^{\infty} \rd k_{\|} \; k_{\|} I_{\perp}
\label{Sparaperp}
\end{equation}
and, in turn,
 \begin{eqnarray}
&& \hspace*{-6mm} I_{\|} = \int_{\mathcal C} \rd k_z
   \frac{\omega}{E_{ji}+ \omega} \; R_{TE} (k_z,\kp) \ e^ {2{\rm
       i}k_z{z_0}} \nonumber \\ && - \int_{\mathcal C} \rd k_z
   \frac{\omega}{E_{ji}+ \omega} \; \frac{k_z^2}{k^2} \; R_{TM}
   (k_z,k_\|) \; e^ {2{\rm i}k_z{z_0}}, \label{I_parallel} \\ 
&& \hspace*{-6mm} I_{\perp} = \int_{\mathcal C} \rd k_z
   \frac{\omega}{E_{ji}+ \omega} \; \frac{k_{\|}^2}{k^2} \; R_{TM}
   (k_z,k_\|) \ e^ {2{\rm i}k_z{z_0}}. \label{I_perpendicular} 
\end{eqnarray}

Before proceeding with the evaluation of Eqs.~(\ref{I_parallel}) and
(\ref{I_perpendicular}), we note that both of them are written in terms of
the position $z_0$ of the atom, measured from the center of the slab.
In practice, one would of course want to know the energy shift of the atom
as a function of its distance to the surface of the slab, $\mathcal Z = z_0
- L/2$. Writing Eqs.~(\ref{I_parallel}) and
(\ref{I_perpendicular}) in terms of the atom-surface distance $\mathcal Z$
gives rise to a phase factor $e^{{\rm i}k_zL}$ which we absorb in the
reflection coefficients (\ref{RandT}) by redefining
\begin{equation}
\widetilde{R}_{\lambda}= r_{\lambda} \frac{1- e^{2ik_{\rm zd}L}} {1
  -r_{\lambda}^2 e^ {2ik_{\rm zd}L}}\;,  
\end{equation} 
so that Eq.~(\ref{I_perpendicular}) turns into
\begin{equation}\label{I_parallel_TE}
I_{\perp} = \int_{\mathcal C} \rd k_z \frac{\omega}{E_{ji}+ \omega} \;
\; \frac{k_{\|}^2}{k^2} \; \widetilde{R}_{TM} \; e^ {2{\rm i}k_z{\cal Z}},
\end{equation}
and similarly for Eq.~(\ref{I_parallel}). Since the photon frequency is
given by $\omega = \sqrt{k_{\|}^2 + k_{z}^2}$, one can identify branch
points in the integrand at $k_z = \pm {\rm i} k_{\|}$. We choose to place
the square-root cuts from $k_z = {\rm i} k_{\|}$ to ${\rm i} \infty$ and
from $k_z = - {\rm i} k_{\|}$ to $-{\rm i} \infty$. Apart from this
square-root cut, the integrand is analytic in the upper half-plane and for
$\mathcal Z >0$ vanishes exponentially on the infinite semicircle in the
upper half-plane. Therefore Cauchy's theorem allows us to deform the
original integration path $\mathcal C$ into a new path $\mathcal C'$ that
goes round the square-root cut from ${\rm i} k_{\|}$ to ${\rm i} \infty$
(see Fig.~\ref{contour}). Identifying the correct sheet on each side of the
cut by demanding that $\omega >0$ on the real $k_z$ axis and re-expressing
$k_z = {\rm i}q$, we can work out the integral along the path $\mathcal C'$
and obtain
\begin{equation}
I_{\perp}= 2 E_{ji} \int_{k_{\|}}^{\infty} \rd q \; \frac{k_{\|}^2} 
{\sqrt{q^2 - k_{\|}^2}} \; \frac{\widetilde{R}_{TM}} {E_{ji}^2 - k_{\|}^2 + 
q^2} \; e^ {-2q{\cal Z}}.
\label{resI} 
\end{equation} 
The calculation of the parallel contribution $I_\parallel$,
Eq.~(\ref{I_parallel}), runs along exactly the same lines. Substituting
Eq.~(\ref{resI}) into Eq.~(\ref{Sparaperp}), we see that the two
contributions $S_{\|}$ and $S_{\perp}$ to the energy shift
(\ref{deltaE-ret}) are both double integrals, over $k_\|$ and over 
$q=-{\rm i}k_z$. We choose to make a change of variables to 
$u = (q^2 - k_{\|}^2)^{1/2}/E_{ji}$  and $v= k_{\|}/E_{ji}$, and arrive at
\begin{eqnarray}
 S_{\|}\!\! &=&\!\!  
\frac{1}{4} \int_{0}^{\infty}\!\! \rd v \int_{0}^{\infty}\!\! \rd u \; 
\frac{v}{\sqrt{u^2 + v^2}} \frac{1} {1 + u^2} \label{S_paral}\\
&&\times \left[(u^2 + v^2) \widetilde{R}_{TM} -u^2 \widetilde{R}_{TE} 
\right] \; e^{-2 \mathcal Z E_{ji} \sqrt{u^2 + v^2}},  \nonumber 
\end{eqnarray}
and
\begin{eqnarray}
S_{\perp} \!\! &=&\!\! \frac{1}{2}  
\int_{0}^{\infty}\!\! \rd v \int_{0}^{\infty} 
\!\! \rd u \; \frac{v^3}{\sqrt{u^2 + v^2}} \; \frac{1} {1 + u^2}  \; 
\widetilde{R}_{TM} \nonumber\\
&&\times\  e^ {- 2 \mathcal Z E_{ji} \sqrt{u^2 + v^2}}.
\label{S_perp}
\end{eqnarray}
These expressions, together
with Eq.~(\ref{deltaE-ret}), give us a general formula for the energy shift
of a ground-state atom in front of a dielectric slab. The shift depends on
the matrix elements of the atomic dipole between the initial state
$|i\rangle$ and other states $|j\rangle$ that are coupled to $|i\rangle$ by
strong dipole transitions. In practice, the sum in Eq.~(\ref{deltaE-ret})
over intermediate states $|j\rangle$ is in most cases dominated by a single
close-lying state with a strong dipole transition to the initial atomic
state $|i\rangle$. Alternatively, we can use the identity
\begin{equation}\label{mu-p-identity}
\left| \langle j |\boldsymbol{\mu}_{\sigma}\right|i \rangle|^2 = 
\frac{4\pi\varepsilon_0\alpha} {m^2E_{ji}^2} \left| \langle j| 
{\bf p}_{\sigma}|i \rangle \right|^2,
\end{equation}
where $\alpha= e^2/ 4 \pi \varepsilon_0$ is the fine structure 
constant, and re-write the energy shift in terms of the matrix elements of
the momentum operator,
\begin{equation}\label{deltaE}
\delta E = -\frac{2 \alpha}{\pi m^2}
\sum_{j} \sum_{\sigma=\|,\perp} \; E_{ji} S_{\sigma} |p_{\sigma}|^2\;. 
\end{equation}
In this form the shift is very similar to that of an atom in front of a
dielectric half-space (cf. Eqs.~(2.12), (2.25), and (2.26) of
Ref.~\cite{Eberlein}), except for the different reflection coefficients in
each situation.

In order to further analyse or calculate the energy shift numerically, it is
convenient to transform the double integrals in Eqs.~(\ref{S_paral}) and
(\ref{S_perp}) into polar coordinates by substituting $u=s \cos \phi$ and
$v=s \sin \phi$, and then replace $\phi$ by $t= \cos\phi$. This gives
\begin{equation}\label{Spara-polar}
 S_{\|} = \frac{1}{4} \int_{0}^{\infty} \rd s \int_{0}^{1} \rd t 
 \; \frac{s^3} {s^2t^2+ 1} \; \left(\widetilde{R}_{TM} -t^2
\widetilde{R}_{TE}\right)  \; 
e^{- 2 \mathcal Z E_{ji} s},
\end{equation}
\begin{equation}\label{Sperp-polar}
 S_{\perp} = \frac{1}{2} \int_{0}^{\infty} \rd s \int_{0}^{1} \rd t 
 \; \frac{s^3} {s^2t^2+ 1} \; \left(1-t^2\right)\widetilde{R}_{TM} \; 
e^{- 2 \mathcal Z E_{ji} s}, 
\end{equation}
 with reflection coefficients
 \begin{eqnarray} \label{R_TE-t2}
 && \hspace*{-5mm} \widetilde{R}_{TE} = \frac{-(n^2-1)t^2} {2 + (n^2-1) 
 t^2 + 2 \sqrt{1+ (n^2-1) t^2} \coth \Lambda}
 \\ \label{R_TM-t2} &&  \hspace*{-5mm} \widetilde{R}_{TM} 
= \frac{n^4-1-(n^2-1) 
 t^2} { n^4 +1 + (n^2-1) t^2 + 2n^2 \sqrt{1+ (n^2-1) t^2} \coth \Lambda} 
\end{eqnarray}
and the abbreviation $\Lambda =LE_{ji}s\sqrt{1+ (n^2-1) t^2})$. 
Thus the energy shift of the atom in front of a dielectric slab is given by
Eqs.~(\ref{deltaE-ret}) or (\ref{deltaE}), with 
Eqs.~(\ref{Spara-polar})--(\ref{R_TM-t2}). In this form the shift is readily
computed numerically, as we shall do in Section \ref{Sec:summary}. However, to
extract important physics and be in the position to make quick estimates,
one should investigate the asymptotic behaviour of the shift in various
physically significant regimes, which we shall do first.

\section{Asymptotic analysis} 
The nature of the interaction of the atom with the slab depends on the
separation between them: for large separations the interaction is manifestly
retarded, but for small separations the retardation can be neglected and the
interaction can be assumed to take place instantaneously. The scale on which
one makes this distinction of the atom-surface separation being small or
large, comes from comparing the time $2 \mathcal Z/c$ that a virtual photon
takes for a round-trip between atom and surface to the time scale of
internal evolution of the atom. For the atom in state $|i\rangle$ with a
strong dipole transition into a close-lying state $|j\rangle$, the time
scale of the atom's internal dynamics is given by $\hbar/E_{ji}$.  The ratio
of the two time scales is $2 \mathcal Z E_{ji}$ in natural units, which can
therefore be used as the criterion for retardation: the interaction is
manifestly retarded for $2 \mathcal Z E_{ji}\gg 1$, as the atom has evolved
appreciably by the time the virtual photon has completed its round-trip, and
it can be considered non-retarded for $2 \mathcal Z E_{ji}\ll 1$, because
the atomic state hardly changes while the photon travels to the surface and
back. In terms of length scales, it is the relative sizes of the distance
$\mathcal Z$ of the atom from the surface and the wavelength $1/E_{ji}$ of
the strongest internal transition that matter.  However, the thickness of
the slab $L$ provides a third length scale to consider. We shall now
consider the various asymptotics limits.

\subsection{Thick slab $L\gg \mathcal Z$}
For a very thick slab, i.e. in the limit $L\rightarrow\infty$, we can
approximate $\coth \Lambda\simeq 1$ in Eqs.~(\ref{R_TE-t2}) and
(\ref{R_TM-t2}). Then expressions (\ref{Spara-polar}) and
(\ref{Sperp-polar}) reduce to what they would be for a dielectric half-space
\cite{Eberlein}. The energy shift for an atom in front of a dielectric
half-space has been analysed in detail previously in both the retarded and
the non-retarded limits \cite{Wu-paper1}.

\subsection{Thin slab $L\ll \mathcal Z$}
If the atom-surface separation $\mathcal Z$ is much larger than the slab
thickness $L$ then the exponentials in Eqs.~(\ref{Spara-polar}) and
(\ref{Sperp-polar}) effectively cut off the integral at very small values of
$s$, so that the argument of the $\coth$ in Eqs.~(\ref{R_TE-t2}) and
(\ref{R_TM-t2}) stays very small throughout the whole effective range of
integration. Thus we can approximate the $\coth$ by its small-argument
expansion, $\coth \Lambda\simeq 1/\Lambda$. This leads to significant
simplifications in Eqs.~(\ref{R_TE-t2}) and (\ref{R_TM-t2}) because the
square roots in the denominators drop out, and we get:
 \begin{eqnarray} \label{R_TEsmallL}
 && \hspace*{-5mm} \widetilde{R}_{TE} \simeq \frac{-(n^2-1)t^2} {2 + (n^2-1) 
 t^2 + 2/(sLE_{ji}) }
 \\ \label{R_TMsmallL} &&  \hspace*{-5mm} \widetilde{R}_{TM} 
\simeq \frac{n^4-1-(n^2-1) 
 t^2} { n^4 +1 + (n^2-1) t^2 + 2n^2/(sLE_{ji})}\;. 
\end{eqnarray}
Substituting these into Eqs.~(\ref{Spara-polar}) and (\ref{Sperp-polar}), we
can now carry out the $t$ integral. While elementary, this integration gives
an unwieldy combination of rational functions, square roots, and $\arctan$,
so that we dispense with writing it down. The subsequent integration over
$s$ cannot be performed analytically, unless we make further approximations,
which we shall do in the following for the retarded and non-retarded limits.

\subsection{Retarded regime ($2 \mathcal Z E_{ji}\gg 1$) for a ``thin'' slab}
If $2 \mathcal Z E_{ji}\gg 1$ then we can apply Watson's lemma to the $s$
integrals in Eqs.~(\ref{Spara-polar}) and (\ref{Sperp-polar}).  So, we
substitute the approximated reflection coefficients (\ref{R_TEsmallL}) and
(\ref{R_TMsmallL}), carry out the $t$ integration, and then expand the
integrand of the $s$ integral around $s=0$, after which the $s$ integral
over the leading term becomes elementary. In such a way we find
that the energy-level shift is given by
\begin{equation} \label{Force-Lsmall}
\delta E \simeq -\frac{(n^2-1) L}{160 \pi^2\varepsilon_0 n^2 \mathcal Z^5} 
\sum_{j \neq i} \; \frac{(5+ 9n^2)|\mu_{\|}|^2 + 2 (4+ 5n^2)|\mu_{\perp}|^2 }
    {E_{ji}}.  
\end{equation}
We would like to note that this result is valid for $2 \mathcal Z E_{ji}\gg
1$ and $\mathcal Z\gg L$, but other than that for any slab thickness $L$. In
particular, there is no restriction on $LE_{ji}$, which can have any size
$>$ or $<1$ provided it is much smaller than $\mathcal ZE_{ji}$. In this
sense the notion of a ``thin'' slab is slightly misleading in the retarded
regime, as any slab of finite thickness can be considered thin for large
enough $\mathcal Z$.

Another interesting aspect of this result is that it shows that there is
absolutely nothing unusual or non-analytic about the limit $L\rightarrow
0$. A calculation of the Casimir-Polder force using field-theoretical means
and four photon polarizations in a Gupta-Bleuler quantization scheme
\cite{Bordag-sheet} has found different results for different ways of
implementing the boundary conditions on the photon field, and the tentative
explanation for this discrepancy, as given in Ref.~\cite{Bordag-sheet}, has
been that these different results should apply to thick and thin
slabs. However, this explanation is inconsistent with our explicit results
for the Casimir-Polder energy shift for slabs of arbitrary finite thickness.

\subsection{Non-retarded regime ($2 \mathcal Z E_{ji}\ll 1$)}
In the non-retarded limit the interactions between the atom and the slab can
be approximated as instantaneous, and the energy shift in this regime can be
calculated by considering the limit $E_{ji}\rightarrow 0$. One could take
this limit in Eqs.~(\ref{Spara-polar})--(\ref{R_TM-t2}), with
Eqs.~(\ref{deltaE-ret}) or (\ref{deltaE}), but the calculation is much
shorter if instead we go back to
Eqs.~(\ref{Sparaperp})--(\ref{I_perpendicular}), which was before we had
deformed the contour $\mathcal C$ in the complex $k_z$ plane. In the limit
$E_{ji}\rightarrow 0$, we get $\omega/(E_{ji}+\omega)\rightarrow 1$ in
Eqs.~(\ref{I_parallel}) and (\ref{I_perpendicular}), so that the square root
cut due to $\omega=\sqrt{k_\|^2+k_z^2}$ disappears from them. Instead, we get
poles at $k_z=\pm {\rm i}k_\|$. When we close the contour $\mathcal C$ in
the upper half-plane, we pick up the residues of the integrands at $k_z=
{\rm i}k_\|$, so that Eqs.~(\ref{I_parallel}) and (\ref{I_perpendicular})
turn into
\begin{eqnarray}
I_\|&=& - \left. 2\pi{\rm i}\; \mbox{Res} \frac{k_z^2\;
\widetilde{R}_{TM}(k_z,k_\|)}{(k_z-{\rm i}k_\|)
(k_z+{\rm i}k_\|)}\; e^{2{\rm i}k_z{\mathcal Z}}
\right|_{k_z={\rm i}k_\|}\\
I_\perp &=& \left. 2\pi{\rm i}\; \mbox{Res} \frac{k_\|^2\;
\widetilde{R}_{TM}(k_z,k_\|)}{(k_z-{\rm i}k_\|)
(k_z+{\rm i}k_\|)}\; e^{2{\rm i}k_z{\mathcal Z}}
\right|_{k_z={\rm i}k_\|}\;,
\end{eqnarray}
which are straightforward to determine. Substituting the results into
Eqs.~(\ref{Sparaperp}) and (\ref{deltaE-ret}), we obtain for the energy
shift in the non-retarded regime 
\begin{eqnarray}
&& \hspace*{-10mm} \Delta E_{{\rm es}}= -\frac{1}{16\pi\varepsilon_0} \;
  \frac{n^2 -1} {n^2 +1} \sum_{j\neq i} \left( 2 |\mu_{\perp}|^2 +
  |\mu_{\|}|^2 \right) \nonumber \\ && 
\times \int_{0}^{\infty} \rd k \; k^2 \; e^{-2 \mathcal Z k} \frac{1-
  e^{-2kL}}{1- \left( \frac{n^2 -1}{n^2 +1} \right)^2
  e^{-2kL}}. \label{Ees-alternative}
\end{eqnarray}
The same result could be achieved from purely electrostatic
considerations. The atom can be viewed as a dipole, and the dielectric slab
can be modelled as a series of image dipoles. The energy shift is then just
the Coulomb interaction energy of the atomic dipole and its images. We show
in Appendix \ref{app:es} that an electrostatic calculation of this sort indeed
reproduces the energy shift (\ref{Ees-alternative}).

Finally we consider the limit $L\ll\mathcal Z$ and obtain for the
non-retarded energy shift of an atom near a thin slab
\begin{equation}
\Delta E_{{\rm es}}\simeq -\frac{3(n^4-1)}{256\pi\varepsilon_0 n^2}
\frac{L}{\mathcal Z^4}
\sum_{j\neq i} \left( 2 |\mu_{\perp}|^2 +
  |\mu_{\|}|^2 \right)\;.\label{Ees-thin}
\end{equation}

\section{Summary and conclusions}\label{Sec:summary}
We have obtained a general formula for the energy-level shift in a
ground-state atom near a non-dispersive dielectric slab of refractive index
$n$: Eq.~(\ref{deltaE-ret}), or alternatively Eq.~(\ref{deltaE}), with the
parallel and perpendicular contributions $S_{\|}$ and $S_{\perp}$ given by
(\ref{Spara-polar}) and (\ref{Sperp-polar}), respectively. While given only
as a double integral, it is nevertheless in a form that is readily amenable
to both numerical calculations and analytic approximations. We have given
appropriate asymptotic formulae in both the retarded ($2 \mathcal Z E_{ji}
\gg 1$) and the non-retarded regimes ($2 \mathcal Z E_{ji} \ll 1$). For the
latter we showed that the result can be reproduced by means of a classical
electrostatic treatment. For thin slabs the electrostatic energy shift
varies as $L/ \mathcal Z^4$, as shown in Eq.~(\ref{Ees-thin}).

In the retarded regime, on the other hand, our general formula reduces to
Eq.~(\ref{Force-Lsmall}), showing that the shift behaves as $L/ \mathcal
Z^5$, provided $L\ll \mathcal Z$. For this case, it is possible to compare
our result with the one given in Eq. (218) of Ref.~\cite{Buhmann} for the
interaction energy between a ground-state atom and a magnetodielectric
plate,
\begin{equation}\label{BuhmannEq}
U(\mathcal Z) = -\frac{\hbar c \alpha(0)}{160 \pi^2 \varepsilon_0}
\frac{L} {\mathcal Z^5} \left [ \frac{14 \varepsilon^2(0)-
    9}{\varepsilon(0)} - \frac{6 \mu^2(0)- 1} {\mu(0)} \right ],
\end{equation}
where $\alpha(0)$ is the static polarizability of the atom. In order to
compare this result to our result (\ref{Force-Lsmall}), we need to
substitute $\varepsilon(0)= n^2$ for the static dielectric constant and
$\mu(0) =1$ for the static magnetic permeability. Furthermore, the diagonal
elements of the atomic polarizability are
\begin{equation}
\alpha_{\nu\nu}(\omega)=\sum_j \frac{2E_{ji}\left|
\langle j|\mu_\nu|i\rangle\right|^2}{E_{ji}^2-\omega^2}\;,
\quad \nu=\{x,y,z\}\;,
\end{equation}
so that we get for the static polarizability of the isotropic atom
considered in Ref.~\cite{Buhmann}
\begin{equation}
\alpha(0) = 2 \sum_{j\neq i} \frac{|\mu_\nu|^2}{E_{ji}}\;,
\quad \nu=\{x,y,z\}\;.
\end{equation}
In this language our expression (\ref{Force-Lsmall}) reads
\begin{equation} 
\delta E = -\frac{\alpha(0)}{160 \pi^2 n^2 \varepsilon_0} \frac{L}{\mathcal
  Z^5}(n^2-1)(9 + 14n^2),
\end{equation}
which agrees with Eq.~(\ref{BuhmannEq}) upon substitution of
$\varepsilon(0)= n^2$ and $\mu(0) =1$.

The great advantage of our general formulae (\ref{deltaE-ret}),
(\ref{Spara-polar}) and (\ref{Sperp-polar}) is that they make it possible to
know how the energy shift behaves for various slab thicknesses and values of
the atom-surface separation $\mathcal Z$.  Using these formulae and standard
software packages such as Mathematica or Maple, one can easily plot $\delta
E$ for any desired parameter ranges.  In order to plot some examples in a
meaningful and informative way, we
re-write the energy shift in the following form
\begin{equation}\label{shiftSLAB-norm}
\delta E = -\frac{1}{4 \pi \varepsilon_0}
\sum_{j \neq i} \; \frac{1}{4 \pi E_{ji} \mathcal Z^4} \left( W_{\|}^
{\rm slab} |\mu_{\|}|^2 + W_{z}^{\rm slab} |\mu_{z}|^2 \right),
\end{equation}
with parallel part and perpendicular contributions defined by
\begin{equation}\label{W-functions}
 W_{\|}^{\rm slab} = 64 {\mathcal Z}^4 E_{ji}^4 S_{\|} \quad {\rm and} \quad
 W_{z}^{\rm slab} = 64 {\mathcal Z}^4 E_{ji}^4 S_{\perp},
\end{equation}
and the functions $S_{\|,\perp}$ given as before in Eqs.~(\ref{Spara-polar})
and (\ref{Sperp-polar}). The motivation for this choice is that {\em (i)}
$W_{\|}$ and $W_{\perp}$ are dimensionless quantities, and {\em (ii)} they
facilitate easy comparison to the standard Casimir-Polder result \cite{CP}
as $W_{\|}= 1= W_{\perp}$ for the retarded energy shift of an atom in front
of a perfect mirror \cite{fn2}. When interpreting the plots it is important
to bear in mind that one needs to multiply with a factor $-1/{\mathcal Z}^4$
in order to judge the distance dependence of the energy shift. For example,
the functions $W_{\|,z}$ (\ref{W-functions}) are linear for small $\mathcal
Z$, showing that the energy shift for small distances behaves as
$-1/{\mathcal Z}^{3}$, as one expects for an electrostatic interaction.

\begin{figure}
\begin{center}
\includegraphics[totalheight=0.3\textheight]{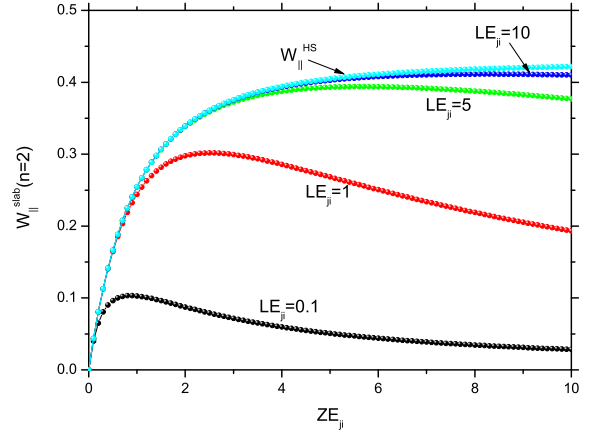} 
\caption{\footnotesize{The function $W_{\|}^{\rm
      slab}$ for various thicknesses of the dielectric slab, with refractive
      index $n=2$.
      The uppermost curve is the result for a dielectric half-space,
      $W_\|^{\rm HS}$. }}\label{Wparal_n2plot}   
\end{center}
\end{figure}

\begin{figure}
\begin{center}
\includegraphics[totalheight=0.3\textheight]{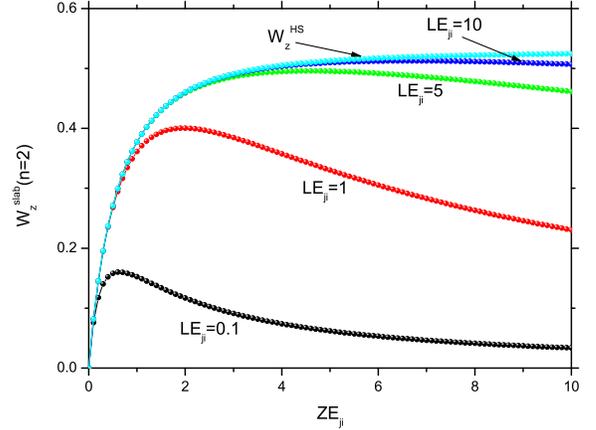} 
\caption{\footnotesize{The function $W_{z}^{\rm
      slab}$ for various thicknesses of the slab, with refractive index
      $n=2$. The uppermost curve is for a dielectric
      half-space, i.e. for $LE_{ji}\rightarrow\infty$.}} 
\label{Wperp_n2plot}  
\end{center}
\end{figure}

In Figs.~\ref{Wparal_n2plot} and \ref{Wperp_n2plot} we have plotted
$W_{\|,z}$ as functions of $\mathcal Z E_{ji}$ for several slab thicknesses
$LE_{ji}$, while fixing the refractive index to $n=2$. We have also included
these functions for the dielectric half-space \cite{Wu-paper1,Eberlein}, which
corresponds to the limit $LE_{ji}\rightarrow\infty$. 
In Fig.~\ref{Wparal_n2plot} we show how the shift varies for different
refractive indices if we fix the thickness of the slab at $LE_{ji}=1$.  In
practice values of $LE_{ji}\sim 10$ might be more realistic, but for those
the energy shift is almost indistinguishable from the one for a dielectric
half-space, as evident from Figs.~\ref{Wparal_n2plot} and \ref{Wperp_n2plot}. 

\begin{figure}
\begin{center}
\includegraphics[totalheight=0.3\textheight]{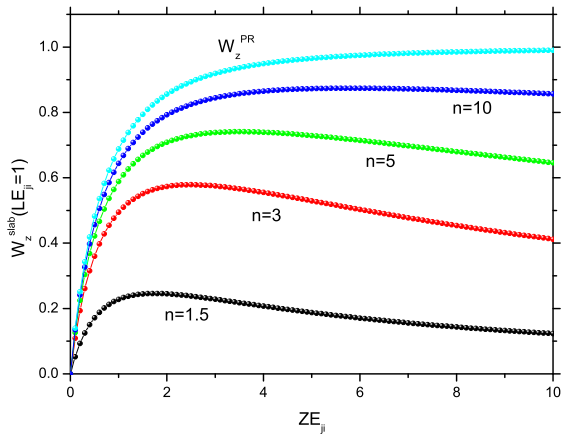} 
\caption{ \footnotesize{The function $W_{z}^{\rm
      slab}$ for a slab of thickness $LE_{ji}=1$ and various 
      values of the refractive index, $n=1.5, 3, 5, 10$. The uppermost
      curve $W_z^{\rm PR}$ is the result for a perfect reflector, i.e. for $n
      \rightarrow \infty$.}}\label{WperpL1plot} 
\end{center}
\end{figure}

\begin{figure}
\begin{center}
\includegraphics[totalheight=0.3\textheight]{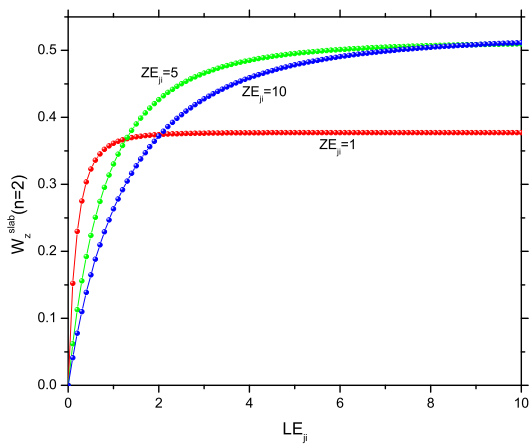} 
\caption{ \footnotesize{The function $W_{z}^{\rm slab}$ as a function of
      $LE_{ji}$, for an atom located at various fixed distances $\mathcal
      ZE_{ji}= 1,5,10$ from the surface. The slab has a refractive index of
      $n=2$.}} \label{Wperp-n2-fLplot}
\end{center}
\end{figure}

\begin{figure}
\begin{center}
\includegraphics[totalheight=0.3\textheight]{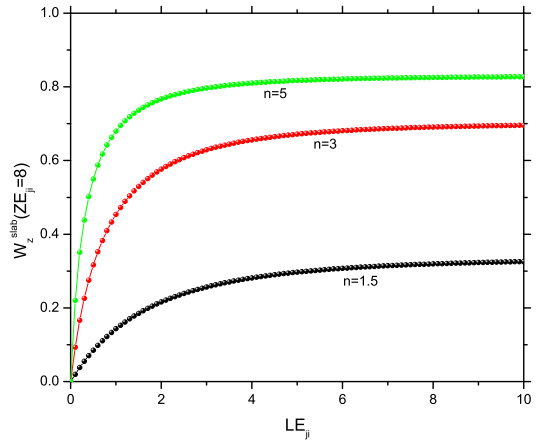} 
\caption{ \footnotesize{The function $W_{z}^{\rm slab}$ as a
     function of $LE_{ji}$, for an atom located at a distance
     $\mathcal ZE_{ji}= 8$ from the slab and various values for the
     refractive index $n=1.5, 3, 5$.} In the limit $n\rightarrow\infty$ this
     function approaches a unit step function.} 
\label{WperpZ8plot}  
\end{center}
\end{figure}

Fig.~\ref{WperpL1plot} shows $W_{z}$ for various refractive
indices $n=1.5, 3,5,10$, and we have also included the limit of a perfect
reflector, $n \rightarrow \infty$, labeled as $W_z^{\rm PR}$.
Furthermore, one can see how the energy shift varies with the thickness of
the slab $LE_{ji}$. In Fig. \ref{Wperp-n2-fLplot}, we have plotted $W_{z}$
as a function of slab thickness for various fixed surface-atom separations,
fixing the refractive index at $n=2$.
For Fig.~\ref{WperpZ8plot} we have fixed the atom's position at $\mathcal
ZE_{ji}= 8$ in the retarded regime, and shown how $W_{z}^{\rm slab}$ varies
with the slab thickness for various values of the refractive index $n$. This
shows again that the retarded Casimir-Polder force has a well-defined and
analytic limit for $L\rightarrow 0$.

\appendix
\section{Mode functions}\label{app:modes}
Throughout this paper we have adopted the same notation as in 
\cite{completeness}, where $\bk^{\pm}$ is the wave vector in vacuum
\begin{equation}\label{bk}
\bk^{\pm} =(k_x,k_y, \pm k_z)= (\bk_{\parallel}, \pm k_z),
\end{equation}
and $\bk_{\rd}^{\pm} = (\bk_{\parallel}, \pm k_{\rm zd})$ is the 
wave vector inside the dielectric slab. The $z$-components of the  
wave vectors in free-space and dielectric are related through 
Snell's law by  
\begin{equation}
k_{\rm zd}= \sqrt{(n^2 -1) k^2_{\|} + n^2 k_z^2},
\end{equation}
and in reverse
\begin{equation}\label{kzd-k}
k_z= \frac{1}{n}\sqrt{k_{\rm zd}^2- (n^2 -1) k^2_{\parallel}},
\end{equation}
which are always positive.

The vector mode functions can be written as a product of a polarization
vector and a scalar mode function,
\begin{equation}
{\bf f}_{\bk \lambda}(\br)= \hat{\bf e}_\lambda\, f_{\bk \lambda}(\br)
\label{scalarf}
\end{equation}
We work with the transverse electric (TE) mode, for which the electric field
is perpendicular to the plane of incidence,
\begin{equation}\label{polvecN}
\hat{\bf e}_{\rm TE} = (-\Delta_{\parallel})^{-1/2}
(-i \partial_y, i \partial_x, 0)\;,
\end{equation}
and the transverse magnetic (TM) mode, for which the magnetic field is
perpendicular to the plane of incidence,
\begin{equation}\label{polvecP}
\hat{\bf e}_{\rm TM} = (\Delta \Delta_{\parallel})^{-1/2}
(-\partial_x \partial_z,
-\partial_y \partial_z, \Delta_{\parallel})\;.
\end{equation}
The momentum space representations $\hat{\bf e}_\lambda(\bk^\pm)$ of the
polarization vectors are obtained by applying the above differential
operators to a plane wave $e^{\rm i\bk^\pm\cdot\br}$.

The scalar mode functions for travelling left-incident modes read
\begin{equation}
f_{\bk \lambda}^{L}({\bf r}) = N \left \{ \begin{array}{ll}
 e^{i{\bf k}^+ \cdot {\bf r}}+ R_{\lambda}  e^{i{\bf k}^- \cdot {\bf
 r}}, & \quad z \leq -L/2 \\ 
 I_{\lambda} e^{i{\bf k}_d^+ \cdot {\bf r}}+ J_{\lambda} e^{i{\bf
 k}_d^- \cdot {\bf r}},  & \quad |z| \leq L/2 \\ 
 T_{\lambda} e^{i{\bf k}^+ \cdot {\bf r}}, & \quad z \geq L/2
 \end{array} \right.  \label{fL}
\end{equation}
for any polarization $\lambda= TE, TM$. The normalization constant is
$N=(2 \pi)^{-3/2}$, and the remaining coefficients are obtained from the 
continuity conditions (\ref{ContCond}); in particular,
\begin{eqnarray}\label{RandT}
R_{\lambda} &=& r_{\lambda} \frac{1- e^{2 {\rm i} k_{\rm zd}L}} {1 -
  r_{\lambda}^2 e^ {2 {\rm i} k_{\rm zd}L}} \ e^{- {\rm i} k_zL} \\
T_{\lambda} &=& \frac{1- r_{\lambda}^2} {1 -r_{\lambda}^2 e^ {2 
{\rm i} k_{\rm zd}L}} \ e^{{\rm i} (k_{\rm zd}-k_z)L}   
\end{eqnarray}
where
\begin{equation}\label{rFresnel}
r_{\rm TE} = \frac{k_z- k_{\rm zd}}{k_z+ k_{\rm zd}} \quad
{\rm and} \quad
r_{\rm TM} = \frac{n^2k_z- k_{\rm zd}}{n^2k_z+ k_{\rm zd}}\;.
\end{equation}

The right-incident modes can be obtained straightforwardly from the 
left-incident modes, by simply inverting the $z$-axis and taking
$z\rightarrow -z$. 
\begin{equation}
f_{\bk \lambda}^{R}(\br) = N \left \{
\begin{array}{ll}
T_{\lambda} e^{i\bk^- \cdot \br} & \quad z \leq -L/2 \\ 
I_{\lambda} e^{i\bk_d^- \cdot \br}+ J_{\lambda} e^{i{\bf
    k}_d^+ \cdot \br} & \quad |z| \leq L/2 \\ 
e^{i\bk^- \cdot \br}+ R_{\lambda}  e^{i\bk^+ \cdot {\bf
    r}} & \quad z \geq L/2\;.
\end{array} \right.  \label{fR}
\end{equation}

The trapped modes are given by
\begin{equation} 
f_{\bk \lambda}^{S,A} ({\bf r}) =  M_{\lambda} \left \{ 
\begin{array}{ll}
 \pm L^{S,A}_{\lambda} e^{{\rm i} \bk_{\|} \cdot \br + \kappa z}, & \quad z 
\leq -L/2 \\ e^{i \bk_d ^+ \cdot \br} \pm e^{i \bk_d ^- \cdot \br},  &  
\quad |z| \leq L/2 \\  L^{S,A}_{\lambda} e^{{\rm i} \bk_{\|} \cdot \br - 
\kappa z}, & \quad z \geq L/2  
\end{array} \right. \label{fSA}
\end{equation}
where the $\pm$ signs apply to the symmetric (S) and antisymmetric (A)
modes, respectively, and $\kappa = |{\rm i} k_z|\geq 0$. Note that for
trapped modes the polarization vector (\ref{polvecP}) is complex and no
longer of unit length. The normalization constants are
\begin{eqnarray}\label{M-TE}
M_{\rm TE} &=& \frac{1}{4\pi \sqrt{ n^2 \frac{L}{2} +
\frac{1}{\kappa} \left(\frac{k_{\parallel}} {k} \right)^2 }} \\ 
M_{\rm TM} &=&  \frac{1} {4 \pi \sqrt{n^2 \frac{L}{2}+ \frac{1}{\kappa}
    \frac{n^2 k_{\|}^2} {k_{\|}^2 + n^2 \kappa^2}}}\;, 
\end{eqnarray}
and by imposing the continuity conditions (\ref{ContCond}) we obtain the 
coefficients $L^{S,A}_{\lambda}$ 
\begin{eqnarray}
L^S_{TE} &=& 2 \cos \left( \frac{ k_{\rm zd} L}{2} \right)
e^{\kappa L/ 2} \label{LNS}\\  
L^A_{TE} &=& 2{\rm i} \sin \left( \frac{ k_{\rm zd} L}{2} \right)
e^{\kappa L/ 2} \label{LNA}  \\
L^S_{TM} &=& 2n \cos \left( \frac{ k_{\rm zd} L}{2} \right)
e^{\kappa L/ 2} \label{L-TM-S}\\  
L^A_{TM} &=& 2 n {\rm i}\sin \left( \frac{ k_{\rm zd} L}{2} \right)
e^{\kappa L/ 2}\;. \label{L-TM-A}  
\end{eqnarray}

The dispersion relations that arise from the simultaneous application of all
matching conditions in Eq.~(\ref{ContCond}) to the symmetric (S) and
antisymmetric (A) modes, with two polarizations $\lambda$ each, read
\begin{equation} 
\kappa = \left \{ 
\begin{array}{ll}
k_{\rm zd} \tan  (k_{\rm zd}L /2 ) & \quad \mbox{for (S), }\lambda=\mbox{TE,} \\
-k_{\rm zd} \cot  (k_{\rm zd}L /2 ) & \quad \textrm{for (A), }\lambda=\mbox{TE,} \\
-k_{\rm zd} \cot  (k_{\rm zd}L /2 )/ n^2 
          & \quad \textrm{for (S), }\lambda=\mbox{TM,} \\
k_{\rm zd} \tan  (k_{\rm zd}L /2 )/ n^2 
          & \quad \textrm{for (A), }\lambda=\mbox{TM,}
\end{array} \right.\label{dispersion}
\end{equation}
where  
\begin{equation}
\kappa= \frac{1}{n}\sqrt{(n^2 -1) k^2_{\parallel}-k_{\rm zd}^2}\;.
\end{equation}

\section{Electrostatic calculation of the electrostatic shift}\label{app:es}
In order to have an independent check of our general formula for the
energy-shift, which in the non-retarded limit takes the form
(\ref{Ees-alternative}), we shall derive the same non-retarded shift purely
by means of a classical electrostatics. If retardation can be ignored, the
energy shift of the atom is simply the electrostatic energy of the atomic
dipole when placed near the dielectric slab. 

If the electrostatic potential $\Phi(r)$ generated by a unit point charge at a
position $\br'$ is known, then the electrostatic energy of an atomic dipole
located at $\br_0$ is (cf. e.g. \cite{cyl} for a more detailed discussion): 
\begin{equation}
\Delta E_{{\rm es}} = \left. \frac{1}{2} \sum_{i=\{x,y,z\}}
\langle \mu_i^2\rangle \mathbf{\nabla}_i
\mathbf{\nabla}'_i \;\Phi_H (\br, \br') \right|_
       {\mbox{\footnotesize ${\bf r} = {\bf r}_0, {\bf r}' = {\bf r}_0$ }}\;. 
\\ \label{DeltaE-es} 
\end{equation}
Here the harmonic function $\Phi_H(\br, \br')$ is the difference between the
potential $\Phi(r)$ generated by the point charge at $\br'$ and the
potential that would be generated by that charge in unbounded space, so as
to exclude from $\Delta E_{{\rm es}}$ the (infinite) electrostatic
self-energies that do not depend on the relative position of the dipole and
the slab. As $\Phi_H(\br, \br')$ is a solution of the Laplace equation and
must vanish for $z\rightarrow\pm\infty$, it must be of the form:
\begin{eqnarray}
&&\hspace*{-5mm}
\Phi_H(\br, \br')= \int_{-\infty}^{\infty} \rd k_x \int_{-\infty}^{\infty} 
\rd k_y \;e^{{\rm i} k_xx+ {\rm i} k_yy}\nonumber\\
&&\hspace*{-5mm}\times\left\{
\begin{array}{ll}
C_1(\bk_\parallel,\br')\; e^{k_\parallel z}&\mbox{for\ }z\leq -L/2 \\
C_2(\bk_\parallel,\br')\;e^{k_\parallel z}+C_3(\bk_\parallel,\br')\; 
e^{-k_\parallel z}&\mbox{for\ }|z|\leq L/2 \\
C_4(\bk_\parallel,\br')\;e^{-k_\parallel z}&\mbox{for\ }z\geq L/2\;. 
\end{array}
\right.\nonumber
\end{eqnarray}
The coefficients $C_{1-4}(\bk_\parallel,\br')$ are easily worked out by
applying the continuity conditions (\ref{ContCond}) to this electrostatic
problem. Straightforward manipulations then give
\begin{eqnarray}
&&  \hspace*{-15mm} \Phi_H = -\frac{1}{4 \pi \varepsilon_0}
  \frac{\varepsilon -1}{\varepsilon +1}  \int_0^{\infty} \rd k \; J_0 (k
  \rho) \; e^{-k (z+ z'-L)}\nonumber \\ 
&& \times  \frac{1- e^{-2kL}} {1- (\frac {\varepsilon-1}{\varepsilon +1})^2
  e^{-2kL}}  \mbox{\ \ \ for\ }z,z'>L/2\;, 
\label{Phi3-tobesolved} 
\end{eqnarray}
with $\rho=\sqrt{(x-x')^2+(y-y')^2}$. It is instructive to rewrite the
denominator as a geometric series
\[
\frac{1}{1-\left(\frac{\varepsilon-1}{\varepsilon+1}\right)^2 e^{-2kL}}
=\sum_{n=0}^{\infty} \left[
\left(\frac{\varepsilon-1}{\varepsilon+1}\right)^2 e^{-2kL} \right]^n
\]
and note that \cite[Eq.~6.611(1.)]{GR}
\[
\int_0^{\infty} \rd k \; J_0 (k\rho)\; e^{-k(z-z')}
= \frac{1}{\sqrt{\rho^2+(z-z')^2}}\;,
\]
which reveals that $\Phi_H(\br, \br')$ can be understood as being due to a
series of image charges generated by repeated reflections between the two
interfaces of the slab \cite{Smythe}. However, expression
(\ref{Phi3-tobesolved}) is more useful for calculations; substituting it
into Eq.~(\ref{DeltaE-es}) gives for the electrostatic energy shift
\begin{eqnarray}
&& \hspace*{-10mm} \Delta E_{{\rm es}}= -\frac{1}{16\pi\varepsilon_0} \;
  \frac{\varepsilon -1} {\varepsilon +1} \sum_j \left( 2 |\mu_{\perp}|^2 +
  |\mu_{\|}|^2 \right) \nonumber \\ && 
\times \int_{0}^{\infty} \rd k \; k^2 \; e^{-2 \mathcal Z k} \frac{1-
  e^{-2kL}}{1- \left( \frac{\varepsilon -1}{\varepsilon +1} \right)^2
  e^{-2kL}}, 
\end{eqnarray}
which, upon replacing $\varepsilon = n^2$, is
in agreement with Eq.~(\ref{Ees-alternative}).

\begin{acknowledgments}
It is a pleasure to thank Robert Zietal for useful comments.
A.M.C.R. would like to acknowledge financial support from CONACYT
M\'{e}xico. 
\end{acknowledgments}


\begin{thebibliography}{99}
\bibitem{fn1} We work in natural units, setting $\hbar=1$ and $c=1$ unless
  explicitly indicated.
\bibitem{fn2}The non-retarded energy shift of an atom in front of a
  perfectly reflecting mirror is given by $W_{\|}= \pi \mathcal ZE_{ji} /4$ and
$W_{z} = \pi \mathcal ZE_{ji} /2$ in Eq.~(\ref{shiftSLAB-norm}).
\end{thebibliography}
\end{document}